\documentclass{appolb}
\usepackage{graphicx}
\usepackage{hyperref}

\newcommand{\eq}{\begin{equation}}
\newcommand{\qe}{\end{equation}}
\newcommand{\eqa}{\begin{eqnarray}}
\newcommand{\qea}{\end{eqnarray}}
\newcommand{\lbar}{{\overline{\ell}}}

\newcommand{\putplot}[1]{
	\includegraphics[width=0.57\textwidth]{pics/#1}
}

\begin{document}
\title{Approaching the chiral point \\ in two-flavour lattice simulations
\thanks{Presented at ``Excited QCD 2014'', Bjela\v{s}nica, February 2--8, 2014.}
}
\author{Stefano Lottini (for the ALPHA Collaboration)
\address{NIC, DESY, Platanenallee 6, 15738 Zeuthen, Germany}}
\maketitle
\begin{abstract}
We investigate the behaviour of the pion decay constant and the pion mass in
two-flavour lattice QCD, with the physical and chiral points as ultimate goal.
Measurements come from the ensembles generated by the CLS initiative using the
$\mathcal{O}(a)$-improved Wilson formulation, with lattice spacing down to about
0.05 fermi and pion masses as low as 190 MeV. The applicability of $SU(2)$
chiral perturbation theory is investigated, and various functional forms,
and their range of validity, are compared. 
The physical scale is set through the kaon decay constant, whose
measurement is enabled by inserting a third, heavier valence strange quark.
\end{abstract}

\begin{flushright}
	\texttt{DESY 14-088}
\end{flushright}

\section{Introduction}
The only well-established, first-principle tool to explore the
low-energy dynamics of QCD without additional assumptions is, to 
date, Lattice QCD; it is the formulation of the theory on a
discrete spacetime that acts as a regulator while opening the way
to numerical computations, thus enabling the study of intrinsically
nonperturbative features of the theory).

On the other hand, there is little doubt that spontaneous breaking of 
the chiral symmetry takes place and that, at least in the limit of
small quark masses, the effective chiral theory known as Chiral Perturbation Theory
($\chi$PT) is a valid description of QCD \cite{Weinberg:1978kz, Gasser:1983yg}:
still, in the former there remain
unknown constants (``low-energy constants'', or LECs) that only the experiment
\textit{sensu lato} (i.e.~including numerical simulations) can constrain
(see, e.g., \cite{Aoki:2013ldr}
for a review of the relation between $\chi$PT and Lattice QCD results).

In this work, data from large-volume two-flavour ($N_\mathrm{f}=2$)
Lattice QCD simulations are
used to test $\chi$PT and determine some of its LECs.
Along the way, different truncations and variations of the predictions
from $\chi$PT are applied and their practical range of validity is examined.
The lattice formulation also introduces possible sources of uncertainty
(e.g.~the system is necessarily finite and discrete), but this seems to be under
control in this investigation.

\subsection{Chiral Perturbation Theory}
$\chi$PT is an effective field theory for the low-energy regime of QCD.
Its fundamental fields are, in the standard formulation, the pion fields
$U$; the theory exhibits spontaneous chiral symmetry breaking by construction,
signaled by the fact that the leading-order LEC $\Sigma$, corresponding
to the chiral condensate, is nonzero. Together with $F$, the pion decay constant
in the chiral limit, those are the two leading-order LECs of the two-flavour
$\chi$PT (the so-called $SU(2)$ $\chi$PT).
The theory is order-by-order renormalisable, implying that there are infinitely many LECs:
for instance, at next-to-leading order there are seven such energy scales $\Lambda_n$,
usually expressed with reference to the scale of the physical pion mass:
$\lbar_n = \log \Lambda_n^2/m_{\pi,\mathrm{phys.}}^2$, for $n=1,\ldots7$.
A distinctive feature of $\chi$PT is that its predictions, usually
expressed as expansions in small quark masses, small pion masses
or analogous, present logarithmic terms (``chiral logarithms'').

Such is the case for the two formulae that 
are needed throughout this work (here we stick to the conventions
in Section 5.1 of \cite{Aoki:2013ldr}): the quark-mass dependence
of the pion mass $m_\pi$ and its decay constant $f_\pi$,
respectively.\footnote{Throughout this work, lowercase symbols $m$ and $f$
refer to masses and decay constants in physical units, while $M=am$ and $F=af$
denote quantities in lattice units, i.e.~in appropriate powers of the lattice spacing.}
After adopting an
independent variable that is as closely related to lattice
measurements as possible, $y = m_\pi^2/(4\pi f_\pi)^2$
(with the physical point at $y_\pi\simeq0.013$),
those two relations, to NNLO, can be written in the form (``$\xi$-expansion''):
\eqa
	M_\pi^2(\beta,y) &=& B^2_\beta \times \label{eq:mpibaseformula}
		\Bigg\{ 1 -\frac12\Big[\lbar_3+2 y_\pi\lbar_4+\log y_\pi\Big] \cdot y +\frac12 \cdot y\log y \\
		& & \qquad+\Big[c_M+\lbar_4+\frac14(\lbar_3+\log y_\pi)^2 +\frac{q_M^2}{90}\Big] \cdot y^2 \nonumber \\
		& & \qquad-\frac12 \Big[ (\lbar_3+\log y_\pi)+\frac{q_M}{6} \Big] \cdot y^2\log y
			+\frac78 \cdot y^2(\log y)^2 \bigg\} \;\;\;; \nonumber \\
	F(\beta,y) &=& F_\beta \times  \label{eq:fpibaseformula}
		\Bigg\{ 1 +\Big[\lbar_4+\log y_\pi+2 y_\pi\lbar_4] \cdot y -1 \cdot y\log y \\
		& & \qquad+\Big[-c_F-2\lbar_4+(\lbar_4+\log y_\pi)^2+\frac{q_F^2}{36}\Big] \cdot y^2 \nonumber \\
		& & \qquad-\Big[ 2(\lbar_4+\log y_\pi)+\frac{q_F}{6} \Big] \cdot y^2\log y+\frac54\cdot
			y^2(\log y)^2 \bigg\} \;\;\;;\nonumber\\
	q_M &=& 60\lbar_{12} -33\lbar_3 -12\lbar_4+52+15\log y_\pi \;\;; \\
	q_F &=& 18\lbar_4 - 15\lbar_{12}+3\log y_\pi - \frac{29}{2} \;\;.
\qea
Here, the overall amplitudes are meant in lattice units, hence they will have different values at each
of values of the inverse bare coupling $\beta$ for which we have
lattice determinations (each $\beta$ is in a one-to-one relation to a lattice spacing).
Two LECs appear here at NLO, namely
$\lbar_3$ and $\lbar_4$; to NNLO, moreover, three more enter, which are $c_F$, $c_M$
and the one associated, similarly as for the other $\lbar_m$,
to the combination $\Lambda_{12}^2=(7\Lambda_1^2+8\Lambda_2^2)/15$,
\cite{Durr:2013goa}.

\section{Lattice computations}
The masses and decay constants were calculated on configurations
generated, within the CLS initiative \cite{cls}, using the
$\mathcal{O}(a)$-improved two-flavour Wilson discretisation
of the QCD action. Three lattice spacings
$a \simeq 0.075, 0.065,$ $0.048$~fm were simulated
in order to get a handle on the continuum limit, corresponding
to $\beta=5.2, 5.3, 5.5$.
All systems satisfy the condition $m_\pi L \geq 4$
(with $L$ the spatial extent of the discretised system), generally
thought to ensure safety from finite-size deviations.
The fifteen available ensembles span a range of pion masses
down to about $192$~MeV,
along which $\chi$PT formulae are fitted.

The pseudoscalar-sector observables $M_\pi, F_\pi$, as well as the quark
mass $M_q$, have been extracted by measurements
of two-point functions of quark bilinears,
which in turn are measured through stochastic sampling,
as detailed in \cite{Fritzsch:2012wq}.
The (bare) quantities are then
known with typically less-than-percent accuracy
(there are renormalisation factors entering afterwards,
which somewhat increase the uncertainties at play.
$Z_A$ alone, in particular, is responsible for about 40\%
of the error on the final decay constants).
Throughout the analysis, errors are propagated carefully
and the methods developed in \cite{Schaefer:2010hu}
are applied in order not to underestimate hidden 
``slow modes'' in the Monte Carlo chains.

For the translation into physical
quantities, scale-setting is a necessary last ingredient.
Usually this is done by measuring a high-precision quantity
on the lattice, say the kaon decay constant $F_\mathrm{K}(\beta)$,
and then using the experimental value $f_\mathrm{K}=155~$MeV to get
$a(\beta) = F_\mathrm{K}/f_\mathrm{K}$ in fermi (note that here $|V_{us}|$ also enters).
This was the procedure followed in this analysis: previous experience 
shows that the kaon decay constant provides
a rather robust scale and allows for less-than-percent precisions on $a(\beta)$.
We refer to \cite{Fritzsch:2012wq} for the technical details on how
valence $s$-quarks are added to a theory with only the $(u,d)$ pair as 
dynamical quark content (\textit{partial quenching}).

\section{Chiral fits to $M_\pi$, $F_\pi$}
In practice, instead of Eq.~\ref{eq:mpibaseformula}, the equivalent
combination $\rho \equiv M_\pi^2/(2 M_q)$
is built and fitted to an analogous chiral formula with just a different
amplitude in front: in the continuum this amplitude would be just $a^3\Sigma/F^2$,
with $\Sigma$ representing
the chiral condensate in physical units.
Moreover, the data are not on the continuum, hence a further term in the overall
amplitude is added modeling the $a^2$ discretisation effects. Finally, for the sake
of convenience, we rewrite this amplitude so as to obtain, directly as a fit parameter,
the adimensional ratio of $\Sigma$ to the physical pion decay
constant cube, $S_0=\Sigma/f_\mathrm{\pi,phys.}^3$.
The NNLO fit function for $\rho$, then, is analogous to Eq.~\ref{eq:mpibaseformula}, but
in place of the overall $B_\beta^2$ we have the prefactor:
\eqa
	\sigma_\beta &=& (S_0+F^2_\beta S_1) F_\beta \cdot \Big\{
			1+y_\pi\lbar_4
			+y_\pi^2(\lbar_4^2+\frac{q_F^2}{36}-c_F) \\
	& & \qquad
			+y_\pi^2\log y_\pi (-\frac16 q_F)
			+\frac14 y_\pi^2(\log y_\pi)^2
		\Big\}^3 \;\;, \nonumber
\qea
the term in curly brackets expressing 
$f_\pi(y_\pi)/f$ as per Eq.~\ref{eq:fpibaseformula} (or variations thereof
when needed). The two datasets, $F_\pi$ and $M_\pi$, are fitted simultaneously,
and for $F_\beta$ we still have three different values, one per each $\beta$.

\begin{figure}[htb]
	\centerline{
		\putplot{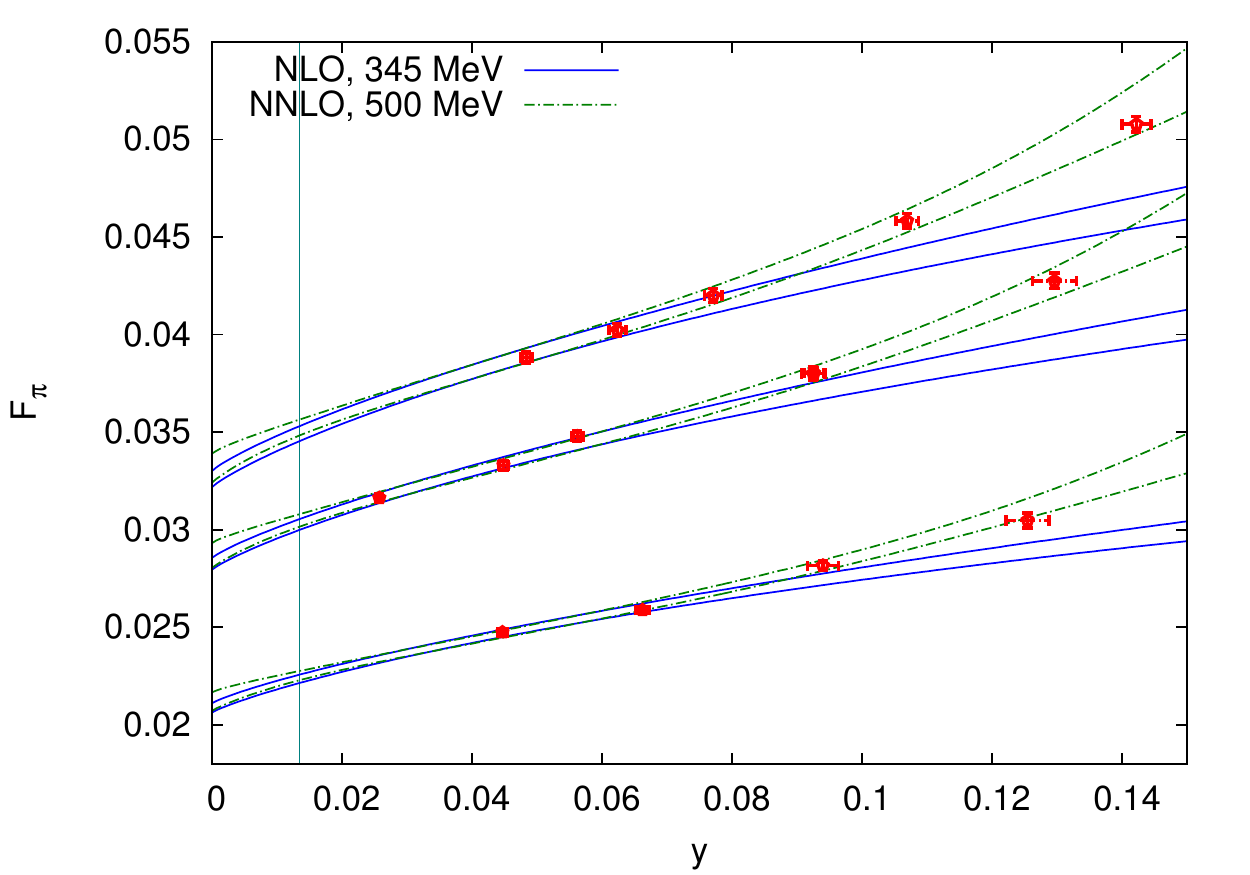}
		\putplot{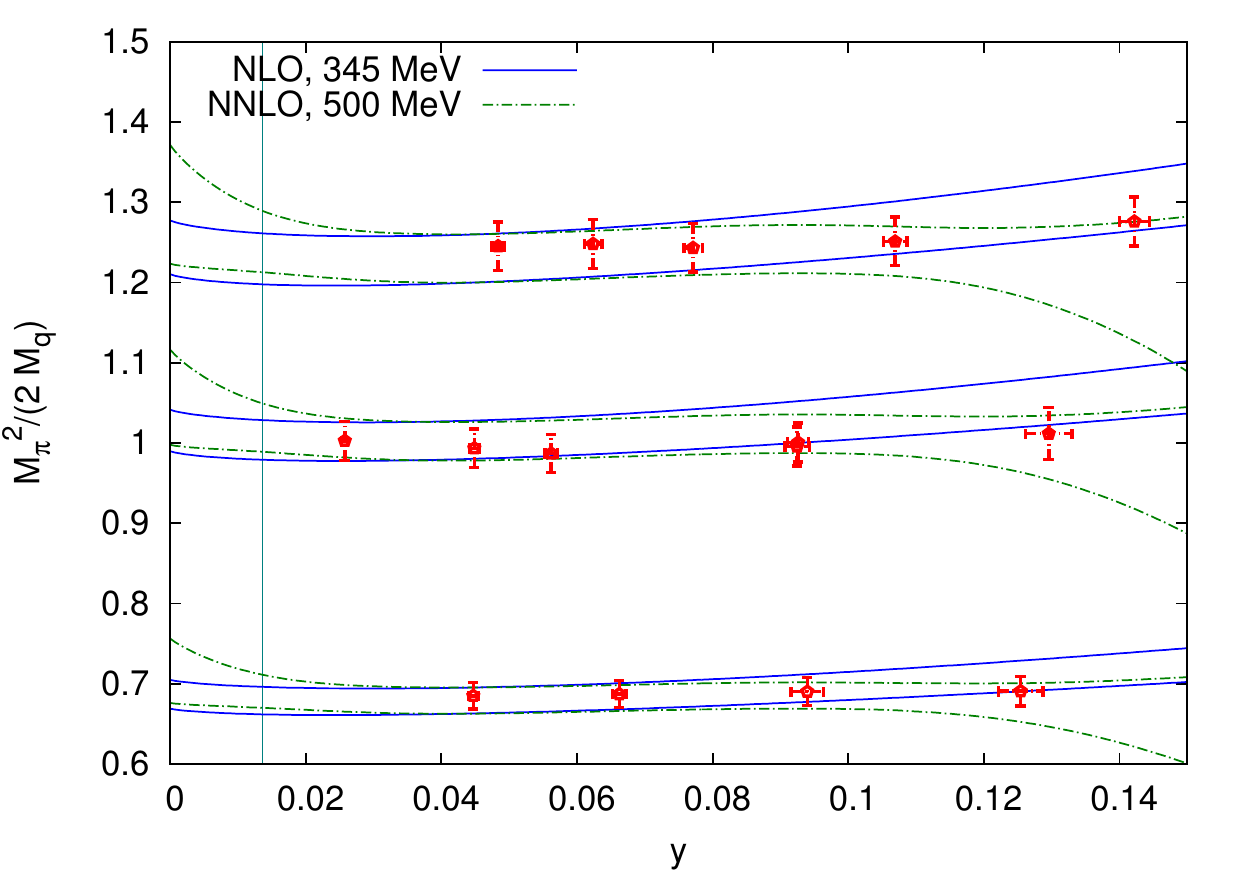}
	}
	\caption{Chiral behaviour of $F_\pi$ (left) and $M_\pi^2/(2M_q)$ (right)
	as a function of $y$.
	Each dataset corresponds to a value of $\beta$ (the topmost line being for the largest
		lattice spacing).
	A one-sigma band is shown
	for the NLO fit with mass-cutoff 345 MeV and for 
	the NNLO cut at 500~MeV. The vertical line marks $y_\pi$.}
	\label{fig:chiralfit_curves}
\end{figure}

Three families of functional forms are attempted: (a) the NNLO formulae shown so far;
(b) their NLO truncation, i.e.~limited to terms of order $y$ and $y\log y$; and (c)
the so-called ``junction'' formulae. The latter come from the empirical observation
that pion observables, in the available range,
seem to just lie on a straight line: on the other hand, we expect $\chi$PT,
close enough to the origin, to take over; hence, we consider a linear function
of $y$ beyond some `junction point' $y>y_\mathrm{jct}$, and the NLO formulae on the left,
with continuity, up to the first derivative, in $y_\mathrm{jct}$.
The total number of fit parameters is 10 for (a) and 7 for (b) and (c).
Also different pion-mass fit ranges are explored, namely
$m_\pi \leq 650, 500, 390$ and $345$ MeV -- corresponding to using 15, 12, 8 and 7
ensembles, respectively (and each ensemble provides \textit{two} data points).

\begin{figure}[htb]
	\centerline{
		\putplot{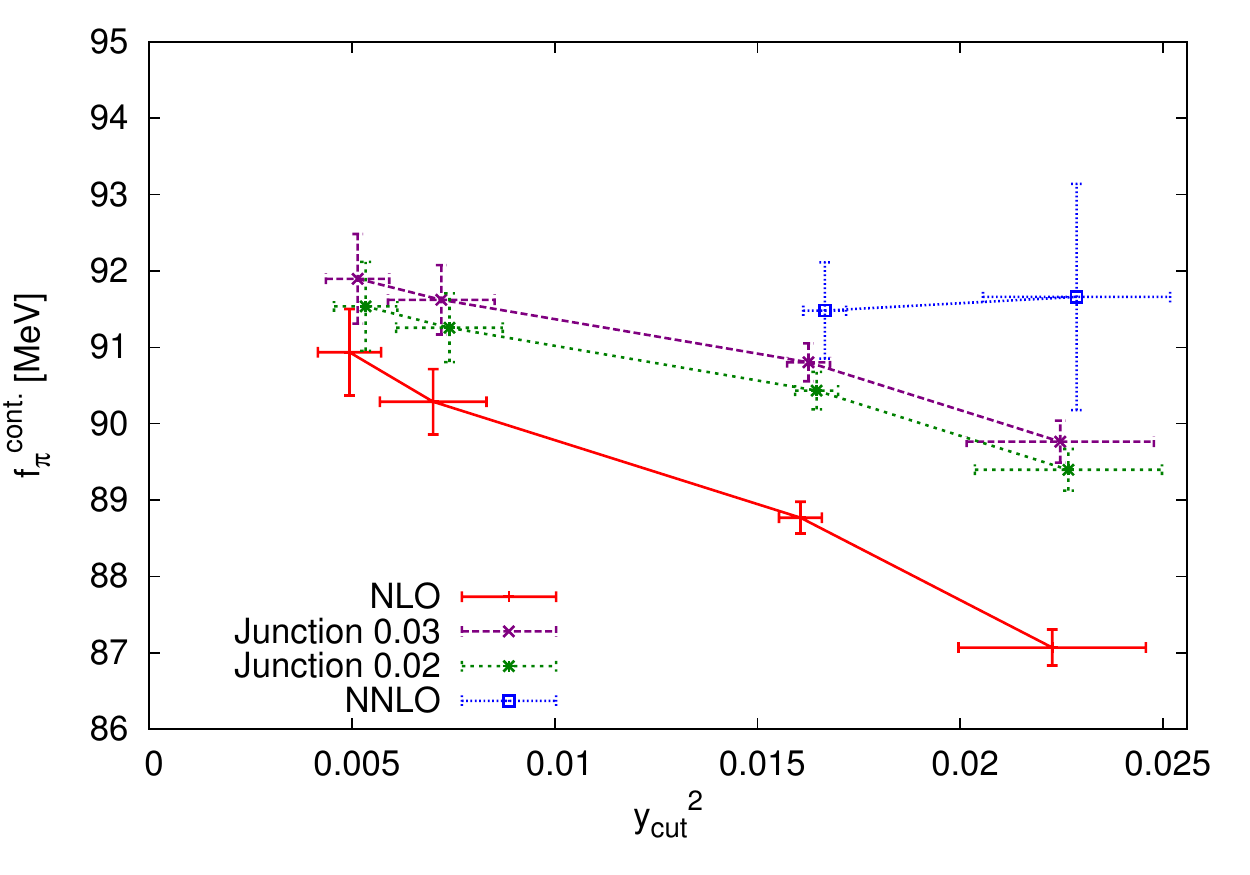}
		\putplot{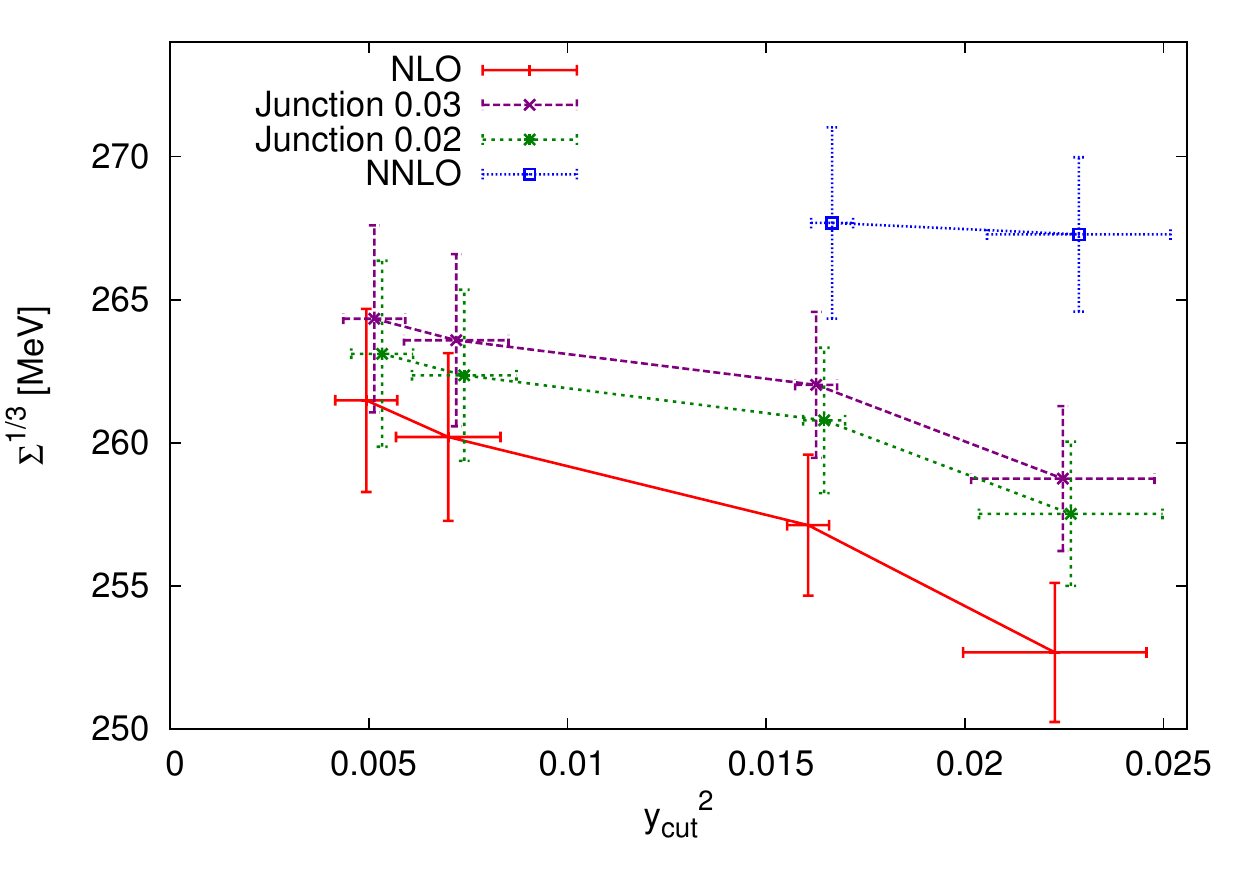}
	}
	\caption{``Stability plots'' for selected fit results: $y_\mathrm{cut}$
	(i.e.~pion-mass-cutoff) dependence
	of $F_{\beta=5.3}$ (left) and of $\Sigma^{1/3}$ (right) for the various fit
	procedures. The horizontal axis is $y_\mathrm{cut}^2$, which should
	roughly make the points lie on a straight line.}
	\label{fig:chiralfit_stability}
\end{figure}

We focus on the results for: $F_\beta$ (and the continuum-limit physical
value $f_\pi$ that stems from it), $\Sigma^{1/3}$, the
higher-order LECs $\lbar_3,\lbar_4,\lbar_{12}$;
the chiral condensate, as well as the quark mass $m_q$, are here always understood to
be expressed in the $\overline{\mathrm{MS}}$ scheme at scale $\mu=2$~GeV.
All fit variants performed will enter an assessment of the systematic
uncertainties involved, generally of higher magnitude than the similar 
analysys performed in the kaon sector. Inspection of ``stability plots''
(i.e.~pion-mass-cutoff dependence of the fit parameters,
see Fig.~\ref{fig:chiralfit_stability}) suggests that as far as $F_\pi$
is concerned NLO with low mass-cut coincides with NNLO, 
hence we trust the former, with the lowest mass-cut,
and take its continuum limit (upon insertion of the $F_\mathrm{K}$-based scale-setting),
which yields $f_\pi = 91.7(1.5)(_{-0.7}^{+0.2})~$MeV.\footnote{The first error is statistical,
	the second is systematic. Note that some conventions keep an additional
	factor $\sqrt{2}$ into $f_\pi$.}
Also the chiral condensate depends weakly on the fit procedure and again we use the NLO
fit and quote $\Sigma^{1/3} = 261.5(3.2)(5.8)~$MeV, a value compatible with the
world-average recently reported in \cite{Aoki:2013ldr}.
A computation of $\Sigma$, based on condensation of Dirac modes near zero,
will be reported in \cite{chiralcondensate}.

As for the higher-order LECs, their precise determination is notoriously
very demanding (see, e.g., \cite{Durr:2013goa} for a recent similar analysis),
in particular due to changes as one adjusts slightly the fit strategy:
we therefore just remark that NNLO fits are unable to give estimates which are
not compatible with zero within their errors, and report the best results
-- again from NLO fits at minimum pion-cut -- for: $\lbar_3=1.3(0.8)$ and $\lbar_4=4.9(0.7)$.
We refer to a forthcoming publication for a more detailed, conclusive
analysis of these $N_\mathrm{f}=2$ data, and in particular an assessment of the
systematic uncertainties on the NLO constants.

\section{Conclusions}
We find that the two-flavour Lattice QCD 
measurements in the pion sector using $\chi$PT-inspired formulae
are much less straighforward to analyse than the analogous ones in the kaon sector
(with the $s$ a valence quark). A general conclusion is that the
range of validity of NLO $\chi$PT (if not of $\chi$PT \textit{tout court})
is shorter, which
prompts the use and comparison of several possible fit functions.
Even so, as soon as one turns her attention to NLO low-energy constants
the uncertainties (especially the systematic ones) are rather large.
In order to provide a thorough characterisation of two-flavour QCD in terms
of its LECs, then, it would be desirable to
include more ensembles with even smaller pion masses;
however, production
of $N_\mathrm{f}=2$ CLS ensembles has been discontinued in favour of $N_\mathrm{f}=2$+$1$ systems,
where presumably similar issues will be seen.

The Author gratefully acknowledges the Conferences organisers for 
their patience and for
providing such an enjoyable and informal atmosphere, as was 
the case in previous editions of Excited QCD as well.

\end{document}